\documentclass[journal,twocolumn]{IEEEtran}

\pdfoutput=1

\usepackage[dvipsnames]{xcolor}
\definecolor{links}{rgb}{0.7,0,0}   %
\definecolor{urls}{rgb}{0,0,0.8}    %
\definecolor{cites}{rgb}{0,0,0.8}   %

\usepackage[colorlinks,hyperindex,linkcolor=links,citecolor=cites,urlcolor=urls]{hyperref} %
 
\usepackage[nosort]{cite}        
\usepackage{url} 
\usepackage[intlimits]{amsmath}
\usepackage{bbm}
\usepackage{graphicx}
\usepackage{paralist}
\usepackage{fancyref}
\usepackage{amssymb}
\usepackage[stretch=16,shrink=16,step=4]{microtype}
\usepackage[font=footnotesize]{caption} 
\usepackage[font=footnotesize]{subcaption}

\DeclareMathOperator{\tr}{tr}

\makeatletter
\def\@IEEEinterspaceratioM{0.265}
\def\@IEEEinterspaceMINratioM{0.1651}
\def\@IEEEinterspaceMAXratioM{0.38}

\def\@IEEEinterspaceratioB{0.31}
\def\@IEEEinterspaceMINratioB{0.19}
\def\@IEEEinterspaceMAXratioB{0.38}
\@IEEEtunefonts
\makeatother
\begin{document}

\IEEEoverridecommandlockouts

\title{Wireless 6G Connectivity for Massive Number of Devices and Critical Services}
\author{%
  Anders~E.~Kal{\o}r,~\IEEEmembership{Member,~IEEE},
  Giuseppe~Durisi,~\IEEEmembership{Senior Member,~IEEE},
  Sinem~Coleri,~\IEEEmembership{Fellow,~IEEE},
  Stefan~Parkvall,~\IEEEmembership{Fellow,~IEEE},
  Wei~Yu,~\IEEEmembership{Fellow,~IEEE},
  Andreas~Mueller,~\IEEEmembership{Member,~IEEE}, and
  Petar~Popovski,~\IEEEmembership{Fellow,~IEEE}
  \thanks{A.~E.~Kal{\o}r is with the Department of Information and Computer Science, Keio
    University, Yokohama 223-8522, Japan, and with the Department of Electronic Systems, Aalborg University, 9220 Aalborg, Denmark (e-mail: aek@keio.jp).

    G.~Durisi is with the Department of Electrical Engineering, Chalmers University of Technology, 412 96 Gothenburg, Sweden (e-mail: durisi@chalmers.se).

    S.~Coleri is with the Department of Electrical and Electronics Engineering, Ko\c{c} University, 34450 Istanbul, Turkey (e-mail: scoleri@ku.edu.tr).

    S.~Parkvall is with Ericsson Research, 164 83 Stockholm, Sweden (e-mail: stefan.parkvall@ericsson.com).

    W.~Yu is with The Edward S. Rogers Sr. Department of Electrical and Computer Engineering, University of Toronto, Toronto, ON M5S 3G4, Canada (e-mail: weiyu@ece.utoronto.ca).

    A.~Mueller is with Robert Bosch GmbH, 70465 Stuttgart, Germany (e-mail: Andreas.Mueller21@de.bosch.com).

    P.~Popovski is with the Department of Electronic Systems, Aalborg University, 9220 Aalborg, Denmark (e-mail: petarp@es.aau.dk).}%
\thanks{The work of A.~E.~Kal{\o}r was supported by the Independent Research Fund Denmark (IRFD) under Grant 1056-00006B.
The work of G.~Durisi was supported in part by the Swedish Research Council under grant 2021-04970 and by the Swedish Foundation for Strategic Research.
The work of S.~Coleri was supported by the Scientific and Technological Research Council of Turkey 2247-A National Leaders Research Grant \#121C314 and Ford Otosan.
The work of W.~Yu was supported by the Natural Sciences and Engineering Research Council (NSERC) via the Canada Research Chairs program and via a Discovery Grant.
The work of P.~Popovski was supported by the Villum Investigator Grant ``WATER'' from the Velux Foundation, Denmark.
}
}%

\maketitle

\begin{abstract}
Compared to the generations up to 4G, whose main focus was on broadband and
coverage aspects, 5G has expanded the scope of wireless cellular systems towards
embracing two new types of connectivity: massive machine-type communication
(mMTC) and ultra-reliable low-latency communications (URLLC). 
This paper discusses the possible evolution of these two types of connectivity within the
umbrella of 6G wireless systems. The paper consists of three parts. The first
part deals with the connectivity for a massive number of devices. 
While mMTC research in 5G predominantly focuses on the problem of uncoordinated access in the uplink for a
large number of devices,
the traffic patterns in 6G  may become more symmetric,
leading to closed-loop massive connectivity. 
One of the drivers for this type of
traffic patterns is  distributed/decentralized learning and inference. 
The second part of the paper discusses the evolution of wireless connectivity
for critical services. While latency and reliability are  tightly coupled in 5G, 
6G will support a variety of safety critical control applications with different types of
timing requirements, as evidenced by the emergence of metrics related to
information freshness and information value. Additionally, ensuring ultra-high reliability for safety critical control applications requires modeling and estimation of the tail statistics of the wireless channel, queue length, and delay. The fulfillment of these stringent requirements calls for the development of novel AI-based techniques, incorporating optimization theory, explainable AI, generative AI and digital twins.
The third part analyzes the coexistence of massive connectivity and critical services. 
Specifically, we consider scenarios in which a massive number of devices need to  support
traffic patterns of mixed criticality. This is followed by a discussion
about the management of wireless resources shared by services with different
criticality. 
\end{abstract}


\section{Introduction}\label{sec:intro}
\begin{figure}
  \centering
    \subcaptionbox{mMTC and URLLC in 5G\label{fig:iot_5g}}[0.45\textwidth]{\includegraphics{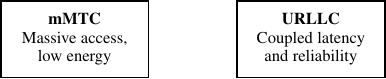}}
    \subcaptionbox{IoT services in 6G\label{fig:iot_6g}}[0.45\textwidth]{\includegraphics{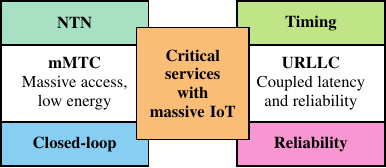}}
    \caption{The evolution of IoT from 5G and 6G. In addition to supporting more devices, mMTC evolves to embrace closed loop, NTN, and to offer critical services with a massive number of devices. URLLC gets augmented to generalized timing and reliability requirements.}\label{fig:iot_5g_6g}
\end{figure}

One of the defining features of 5G was the native support for Internet of Things (IoT) through
massive machine-type communication (mMTC) and ultra-reliable low-latency communications (URLLC).
While mMTC aimed to provide delay-tolerant service for a very large number of low-cost devices,
URLLC targeted human-centric and machine-centric {real-time} applications, such as virtual reality
and self-driving cars. Along with enhanced mobile broadband (eMBB), these services constituted the
``5G triangle,'' which served as the guiding framework for the 5G research and standardization
efforts~\cite{imt2020.m2083-0}. At the conception of 5G, it had been thought that mMTC and URLLC, {as well as their combinations with eMBB}
were sufficient to cover almost all IoT use cases.
Nevertheless, with the evolution and deployment of 5G it started to emerge that the requirements for mMTC and URLLC would need to be redefined, as illustrated in Fig.~\ref{fig:iot_5g_6g}. For instance, the stringent latency requirements of URLLC can be relaxed to a set of more general timing requirements that dynamically adjust to the current state of the system, such as Age of Information (AoI), Age of Loop (AoL), and Value of Information (VoI)~\cite{popovski22timing}. Furthermore, new use cases, especially
centered around machine learning (ML) and artificial intelligence (AI), have introduced new
scenarios that were unthinkable when the 5G vision was initially sketched, such as federated learning and intelligent digital twins. Finally, there is a major change in the overall access infrastructure by embracing nonterrestrial networks (NTNs) \cite{giordani2020non,leyva2020leo}. These networks introduce moving elements in the infrastructure, such as low Earth orbit (LEO) satellites and significantly augment the spatial coverage, thereby the number of potentially accessing devices. Although NTNs will have an important role in massive connectivity, as also indicated in Fig.~\ref{fig:iot_6g}, in this article we will not cover the NTN aspects in detail, as it would require a more elaborate discussion on the specifics of the related network architecture, which is significantly different from the terrestrial one.

As we move towards 6G, the 5G triangle has been expanded to a hexagon, defined by enhanced variants
of the three 5G service types, as well as new scenarios targeting integrated sensing and
communication, integrated AI and communication, and ubiquitous connectivity~\cite{imt2030}. With the
four overarching requirements of sustainability, security/privacy/resilience, ubiquitous
intelligence, and connecting the unconnected, the support for IoT is expected to play an even larger
role in 6G. To fulfill the long-term 6G vision {that involves IoT connectivity} requires, however, progress in three key areas: communication for massive number of devices, communication for critical services, and service coexistence and slicing capabilities. It is important to outline that the denomination ``6G'' can be understood in two different ways. In a broader sense, 6G refers to the future wireless connectivity, accompanied with sensing, positioning, learning, etc. In a narrower sense, 6G refers to the next standardized generation of wireless cellular communication system. The discussion on 6G in this paper should be understood in the broader sense, but in specific parts, such as Section~\ref{sec:standarziation}, we elaborate upon the potential impact on standardization.

The evolution of massive communication from 5G to 6G will be characterized primarily by two aspects.
The first aspect is the even higher device densities targeted in 6G, ranging from $10^6$ to
$10^8$~devices/km\textsuperscript{2}~\cite{imt2030}. This requires more scalable random access
protocols that can resolve transmissions from a large number of uncoordinated, simultaneously
transmitting devices. The second aspect is a widening of the device spectrum, ranging from
zero-energy devices with extremely limited capabilities and long lifetimes, to highly capable
devices equipped with advanced sensors, such as microphones and cameras, and able to execute ML
algorithms locally or in the cloud. By predicting when and what data are relevant at a given time, ML-driven decisions will increasingly govern the traffic patterns and determine the communication targets, since the need to communicate will be triggered by
unpredicted events rather than occurring periodically. For example, an AI may autonomously request data from IoT devices based on what is needed to reduce the uncertainty of a digital twin model, or because it predicts that the data will soon be requested by an external entity. Besides data requests, advanced devices also require frequent firmware updates, and may continuously participate in distributed learning algorithms, such
as federated learning, leading to closed-loop communication. In conclusion, massive connectivity in
6G will shift the focus from uplink-oriented massive IoT to a wide range of traffic patterns
including both small, sporadic transmissions, and closed-loop uplink and downlink driven by interactive applications and distributed learning.

Critical communication refers to communication with strict performance requirements, and was
introduced in 5G as URLLC. The aim was to provide a service that could replace traditional cabled
industrial networks by delivering packets within milliseconds with a reliability in the order of
$1-10^{-5}$ {(five nines) or higher}. While the vision in 6G is to provide even stricter latency and reliability
requirements~\cite{imt2030}, meeting the vision of critical communication requires fundamental
changes. While this paper is focused only on the wireless access part, in a broader perspective, timing and reliability need to be considered \emph{end-to-end} at the system level,
including the mobile core network and various hardware components that influence the overall
reliability. Furthermore, since the initial
vision of 5G was defined, there has been a realization that the latency and reliability can be
decoupled to some extent by focusing on the specific task and goals of the application rather than
the link-level requirements~\cite{popovski22timing}. However, maximizing these benefits requires a
close integration between the communication system and the application itself, and identifying the
control interface will be an important task in the development of 6G. Meeting the complex demands of communication systems, closely linked with control systems and stringent constraints on delay, reliability, availability and resilience, necessitates the development of advanced AI techniques, incorporating optimization theory, explainable AI, generative AI and digital twin, for optimal performance. For instance, AI can help bridging the gap between model accuracy and complexity in interacting systems comprising both wireless channels and physical sensors and actuators, and where the relation between inputs and outputs has no simple mathematical expression. This is particularly important when the tail behavior of the model predictions are crucial, as in systems that require high reliability.

Finally, providing tailored communication services for the diverse range of IoT applications
envisioned in 6G necessitates flexible and efficient resource slicing mechanisms. In 5G, this was
achieved through \emph{network slicing}, {a core network functionality that indicates to the Radio Access Network (RAN) the requirements associated with a specific slice.
This indication is used by the RAN to allocate appropriate radio resources. The most straightforward approach, although far from the most efficient one, is to allocate orthogonal radio (time/frequency/spatial) resources to different slices.
In general, there can be advantages derived from allocating non-orthogonal resources to different slices~\cite{popovski18-12f} and these considerations can be a basis for devising more efficient methods for resource sharing in 6G while keeping the configuration complexity manageable.}

In this article, we present a comprehensive review of the technological developments that have the potential to serve as the foundation for massive and critical connectivity in 6G. The paper is organized as follows. In Section~\ref{sec:massive_connectivity}, we describe the evolution of massive connectivity. We review the massive random access techniques required to support the envisioned device densities, and discuss the role of massive downlink communication. Section~\ref{sec:critical} presents applications for critical services and enabling technologies in 6G. We explore network design for control systems with a broad range of performance metrics that go beyond only latency and reliability, and considerations for ultra-reliable communications, involving finite-blocklength communication and tail statistics related to the channel and queue length. We then present advanced AI based techniques to address the strict constraints in the resulting problems. In Section~\ref{sec:mixed_criticality}, we discuss how to accommodate services with different criticality through flexible network slicing. Concluding remarks are given in Section~\ref{sec:conclusion}.

\section{Massive Connectivity}\label{sec:massive_connectivity}

\subsection{Massive Connectivity in 6G: A Mix of Uplink and Downlink}
\begin{figure}
  \centering
  \includegraphics{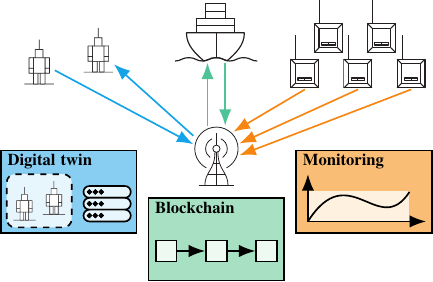}
  \caption{Massive connectivity in 6G will be a mix of uplink and downlink transmissions.}\label{fig:massive_ul_dl}
\end{figure}

Massive connectivity refers to the problem of serving a very large number of users that access the channel in a sporadic fashion, as illustrated in Fig.~\ref{fig:massive_ul_dl}. Traditionally, massive connectivity has been associated with simple IoT devices, such as sensors, that wake up periodically or sporadically to transmit short sensor readings to the base station. This was also the motivation in 5G, where massive connectivity was introduced as the mMTC use case and enabled by the NB-IoT and LTE-M technologies, which provided native support for low-power devices and short packet transmissions through simplified connection establishment procedures~\cite{dahlman20164g}.

Massive connectivity will remain a central component of 6G, where the vision is to support even higher device densities and serving a broader spectrum of device capabilities. In particular, data collection and environmental monitoring will be crucial for important 6G applications, such as AI-based predictions and decision making, which may involve devices that make coordinated, intelligent decisions, such as actively requesting information or initiating model retraining. Some important applications of massive connectivity in 6G include:
\begin{itemize}\setlength\itemsep{1mm}
\item \emph{Sensing and actuation for digital twins:} Digital twinning is being promoted as one of the central use cases in 6G, where physical objects will have virtual representations that can be used for monitoring, simulations, or virtual interactions in, e.g., industrial and smart traffic scenarios~\cite{khan2022digital}. Massive connectivity plays a key role in enabling the data collection that will allow the digital twins to accurately represent their physical counterparts in real-time. The collected data may then be fused with data from other sensors or combined with, e.g., sensing and localization information collected by the base station within the same temporal window~\cite{popovski24time}. Finally, the simulations and interactions with the digital twins may eventually trigger actions or decisions that need to be communicated back to the devices for execution. 
\item \emph{Logistics:} The ubiquitous connectivity of 6G will enable IoT-based worldwide tracking of objects for logistics and supply chain management~\cite{ericsson6Gcyber}. Depending on the complexity of the IoT device, applications may span from simple location reporting to tamper-resistant logging using distributed ledger technology, such as blockchain~\cite{imt2030}.
\item \emph{Massive sensing using zero-energy devices:} Ultra simple, low-cost, battery-less devices that harvest energy from, e.g., vibrations or radio waves drastically simplifies the deployment of IoT. Without the need of a battery, these devices can be made extremely small and printed on things like textiles and warehouse packages for tracking and sensing purposes~\cite{ericsson6Gcyber,Khan_2023}.
\item \emph{Channel access in densely populated areas:} 6G aims to target even higher device densities compared to 5G, such as crowded stadiums~\cite{imt2030}. In such scenarios, allocating dedicated resources to each device for coordination of, e.g., scheduling requests, is impossible, and efficient and scalable massive connectivity technologies are crucial to ensure a good user experience.
\end{itemize}

While the overhead reductions introduced in 5G mMTC were crucial to enable the support of IoT, the access procedures in 5G still perform significantly worse in terms of energy and spectral efficiency compared to what is possible~\cite{agostini24ra6g}. Consequently, 6G will need additional technological innovations to support the envisioned target device densities and applications in 6G without imposing a significant cost in terms of spectral efficiency. In this respect, one central protocol component that can benefit from substantial improvements is the random access procedure, which is needed due to the sporadic uplink transmissions, since the set of transmitting users is unknown a priori to both the transmitters and the base station. Most commercially available mMTC technologies, including those in 5G, are still implemented using techniques that resemble the first random access protocol, namely slotted ALOHA~\cite{roberts75aloha}. While slotted ALOHA protocols are simple to implement, it is well known that they have significant limitations in terms of energy and spectral efficiency, which has also been documented in the context of 5G~\cite{agostini24ra6g}. Furthermore, since random access is central in any massive connectivity protocol regardless of its specific requirements, its performance will be reflected across all applications. Besides random access, several of the new applications envisioned in 6G, such as interactive digital twins, blockchain-enabled logistics, and inference/learning, will bring more symmetric traffic patterns that require equal uplink and downlink, leading to massive closed-loop communication~\cite{ericsson6Gcyber,gunduz23bits}. Since 5G mMTC was primarily targeting uplink traffic~\cite{3gpp_tr38913}, the downlink channel for IoT is another area that will require new innovations in 6G.

In the remainder of this section, we review the enabling technologies for enhancing the massive connectivity support in 6G. We first present an overview of the technologies in the context of grant-based, grant-free, and unsourced random access as the main categories of massive uplink. We then go into details with sparse recovery as a promising and general building block for efficient and scalable uplink connectivity. This is followed by an overview on recent work towards designing massive connectivity for the downlink. We conclude the section by discussing what is needed to integrate these developments into 6G standards and outlining some remaining open problems.

\subsection{Grant-Based, Grant-Free, and Unsourced Random Access for Massive Uplink Connectivity for 6G}\label{sec:grantbased_grantfree}
Broadly speaking, the recent developments in massive random access protocols can be divided into \emph{grant-based}, \emph{grant-free}, and \emph{unsourced} random access. We discuss each category in turn.

\begin{figure}
  \centering
    \subcaptionbox{Grant-based random access\label{fig:grant_based_ra}}[0.45\textwidth]{\includegraphics{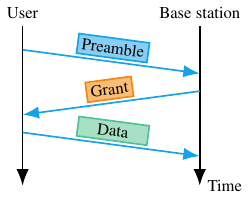}}
    \subcaptionbox{Grant-free random access\label{fig:grant_free_ra}}[0.45\textwidth]{\includegraphics{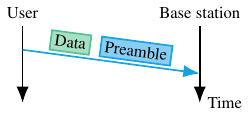}}
    \caption{Grant-based and grant-free random access. In grant-based random access (\subref{fig:grant_based_ra}), the preamble is transmitted first, allowing the base station to assign dedicated resources for the data transmission (i.e., a grant). In grant-free random access (\subref{fig:grant_free_ra}), the preamble and data are transmitted without intermediate feedback from the base station, typically in a sequential fashion.}\label{fig:grant_based_grant_free_ra}
\end{figure}

\subsubsection{Grant-Based Random Access}
Many commercial systems for mMTC, including NB-IoT and LTE-M, rely on grant-based random access. In grant-based random access, the packet transmission contains three steps as illustrated in Fig.~\ref{fig:grant_based_ra}: (1) each active user transmits a random preamble in the uplink; (2) the base station responds with assigned transmission slots (grants) to each of the decoded preambles; (3) each of the granted users transmits its packet in its assigned slot. Due to the dedicated channel resources for the packet transmission in step 3, the packets can be transmitted efficiently provided that the preamble phase, known as activity detection, is successful.

Both NB-IoT and LTE-M use orthogonal preambles in the first step, where each active user transmits a preamble randomly selected from a pool of orthogonal preambles without the possibility of multipacket reception. While this leads to low complexity at the decoder, it is well known that the throughput measured by the fraction of non-colliding transmissions is upper bounded by $1/e\approx 0.37$~\cite{roberts75aloha}. Consequently, the utilization of the channel resources (i.e., the spectral efficiency) is inherently low. Furthermore, since the number of available preambles is limited by the channel coherence length, the number of simultaneously transmitting users that can be served in the same area is also limited. In scenarios with low coherence times, such as the smart traffic scenarios envisioned in 6G, this severely restricts the scalability of the IoT protocols.

The performance of grant-based schemes can be greatly improved by relaxing the orthogonality requirement of the preambles, which significantly increases the number of preambles available within a coherence block. While this can be done without increasing the complexity of the transmitter, it necessitates more sophisticated preamble decoding techniques, such as sparse recovery algorithms that rely on compressed sensing~\cite{chen18sparse,liu18part1}. In this case, the number of available preambles needs to be selected to balance the collision probability, the probability of false alarms, and the complexity of the activity detection algorithm. When the total number of users is not too large, the error is dominated by collisions rather than false alarms, and it may be advantageous to assign a unique preamble to each user to completely avoid collisions. Furthermore, this enables user identification ``for free'', since the transmitting users can be identified based from the decoded preambles. By allocating multiple preambles to each device, it is even possible to embed a few bits of information in the preambles as well, e.g., to indicate the amount of data that the user has to transmit~\cite{senel17embed}. Besides avoiding collisions, the idea of assigning dedicated preambles to each user can be further justified by the result that the probability of activity detection error can theoretically be made to go to zero as the number of base station antennas goes to infinity~\cite{liu18part1}. However, when the total number of users is large, the computational complexity of the activity detection algorithm may be prohibitive, and false alarms may also be more prominent than collisions. In this case, letting the users randomly select their preamble may lead to better overall performance-complexity tradeoff. We return to the use of sparse recovery for massive uplink in Section~\ref{sec:massive_cs}.

The main disadvantage of grant-based random access is the latency and power consumption introduced by the three protocol steps. Furthermore, when the packets are short, the additional overhead introduced by the transmission of the grant message may outweigh the advantage of having dedicated channel resources for the packet transmissions. An alternative strategy that avoids this procedure is grant-free random access, which is discussed next.

\subsubsection{Grant-Free Random Access}
In grant-free random access, illustrated in Fig.~\ref{fig:grant_free_ra}, the active devices transmit their preambles and packets immediately, typically in a sequential manner, without waiting for intermediate feedback from the base station. The recently-added \emph{Small Data Transmission} (SDT) feature in 5G is an example of steps in this direction~\cite{3gpp_ts18321}. To decode the transmitted packets, the base station performs activity detection and channel estimation based on the preambles, and then decodes the messages coherently while treating the interfering transmissions as noise.

As in the case of grant-based random access, grant-free random access suffers
from preamble collisions, and several techniques have been proposed to improve
the performance. One approach is to let the devices transmit multiple
repetitions with specific delays, and then apply successive interference cancellation (SIC) at the decoder, as in, e.g., coded slotted ALOHA~\cite{paolini15tit}. Note
that this comes at the cost of increased power consumption at the devices, as each packet needs to
be transmitted multiple times. Grant-free access can also benefit from non-orthogonal preambles and
sparse recovery for activity detection. Compared to grant-based access, false alarms are
less severe since they do not reduce the spectral efficiency. On the other hand, the estimation error
introduced by the non-orthogonal preambles can significantly degrade the performance of the coherent decoding.
This is further exacerbated by the fact that both preamble and data must be transmitted within a single coherence block, and thus
increasing the preamble length to improve channel estimation reduces the number of symbols available
for data transmission~\cite{liu18part2}. Using ideas from coded slotted ALOHA, this problem can be
mitigated using coded repetitions and SIC, so that the probability of having at least one transmission with a good channel estimation is boosted~\cite{sorensen18codedpilot}.

\subsubsection{Unsourced Random Access}
Unsourced random access refers to an information-theoretic formulation of the grant-free massive access problem proposed in~\cite{polyanskiy17ura}. Rather than treating it as a separate category, it is therefore more meaningful to consider it as a specific instance of grant-free random access. In unsourced random access, the active users encode their messages using a common codebook, and the codewords are transmitted simultaneously over a shared channel. The task of the decoder is to produce an unordered list of the transmitted messages. The fact that all users share the same codebook implies that the messages cannot contain identification information (hence the term ``unsourced''). Initially, this constraint arose as a consequence of the desire to study the limits of random access in the asymptotic regime, where the number of users approaches infinity. Since user identification would then require an infinite number of bits, including user identification would force the transmission rate to zero. However, this is purely an information theoretic implication that, for any practical purpose, can be circumvented by including the user identity in the message itself (i.e., handling the identification problem at a higher layer, although strictly speaking it reduces the transmission scheme back to sourced random access). The main advantage of unsourced random access appears when the number of users is very large or changes frequently, thus keeping track of the preambles assigned to each user is cumbersome, as in, e.g., satellite networks, or in applications where the identities of the transmitters are not important, but only their messages (e.g., a message \emph{fire} could be important regardless of where it comes from).

The majority of the protocols developed for unsourced random access rely on sparse recovery, possibly combined with additional coding~\cite{amalladinne20tit,fengler21tit}, although techniques based on, e.g., tensor decomposition have been proposed as well~\cite{decurninge21wcl}.
Specifically, the work in~\cite{amalladinne20tit} proposes a scheme where each user transmits its message across multiple sub-blocks. Each sub-block is transmitted non-orthogonally with the other active users, and using sparse recovery techniques the receiver tries to recover the transmissions within each sub-block. Using an outer tree code, the receiver tries to stitch together the decoded transmissions from each sub-block into complete messages. The same idea is adopted in~\cite{fengler21tit}, but with an alternative sparse recovery algorithm that is better suited for handling a very large number of preambles.
Finally, we note that many random access schemes designed for grant-free access can also be made unsourced by making the preamble selection random or determined based on the message data (as opposed to assigning a unique preamble to each user).

\subsubsection{Grant-Based vs.\ Grant-Free Massive Random Access}
Both grant-based and grant-free methods can be useful in appropriate setups. Generally speaking, grant-based schemes tend to perform better~\cite{kang22tcom}, especially for larger payloads, but the three phases increase latency and power consumption. Grant-free schemes, on the other hand, tend to suffer from channel estimation errors. Therefore, multiple transmission repetitions are needed to achieve a performance that is comparable to that of grant-based schemes, which in turn also increases latency and power consumption. In conclusion, the grant-based approach appears generally as a more attractive choice. However, grant-free schemes might be a better choice for ultra low-power devices that can tolerate the larger error probability, or for uplink-only devices, such as zero-energy devices. Finally, grant-free might be attractive when the payload size is very small so that the additional overhead incurred by the grant-based approach becomes significant in comparison. Combinations of the two schemes can also be envisioned for 6G. For example, data transmission can start using a grant-free scheme and seamlessly be moved to a grant-based scheme in case the amount of data to transmit is large.

\subsection{Sparse Recovery for Massive Random Access}\label{sec:massive_cs}
As argued in Section~\ref{sec:grantbased_grantfree}, many modern random access methods rely on the use of non-orthogonal preambles to enlarge the set of available preambles, so as to reduce the probability of preamble collisions. However, the non-orthogonality of the transmitted preambles makes the activity detection problem nontrivial. The predominant strategy is to apply sparse recovery techniques that exploit the sparse user transmission patterns that result from the fact that only a small fraction of the users are active at any given time. In this section, we provide a brief review of the foundations behind this strategy in the context of massive random access. To simplify the presentation, we will focus on the case where each user is assigned a unique preamble, which precludes collisions. Clearly, supporting a massive number of users under this strategy is only possible with a very large number of preambles, which necessitates the preambles to be non-orthogonal. We consider a scenario with $N$ single-antenna devices and a single base station with $M$ antennas. Let $\mathbf{s}_k\in\mathbb{C}^{L\times 1}$ denote the preamble assigned to user $k$, $k=1,\ldots,N$, which is assumed to
be known at the base station. Suppose a random number $K$ out of the $N$ users are active, and let $a_k=1$ if user $k$ is active, otherwise $a_k=0$. The channel between device $k$ and the base station is assumed to be given by the random vector $g_k\mathbf{h}_k\in\mathbb{C}^{M\times 1}$, composed by large scale fading coefficient $g_k\in\mathbb{R}$, and unknown Rayleigh fading coefficients $\mathbf{h}_k\in\mathbb{C}^M$ drawn independently from $\mathcal{CN}(0,1)$. The signal $\mathbf{Y}\in\mathbb{C}^{L\times M}$ received by the base station can then be written as
\begin{align}
  \mathbf{Y} &= \sum_{k=1}^N \sqrt{g}_k a_k \mathbf{s}_k \mathbf{h}_k^T  + \mathbf{W}\\
  &=
  \begin{bmatrix}
    \mathbf{s}_1\, \cdots\, \mathbf{s}_N
  \end{bmatrix}
  \begin{bmatrix}
    a_1 \sqrt{g}_1&&\\
    &\ddots&\\
    &&a_N \sqrt{g}_N
  \end{bmatrix}
  \begin{bmatrix}
    \mathbf{h}_1^T\\
    \vdots\\
    \mathbf{h}_N^T
  \end{bmatrix}
  + \mathbf{W}\\
  &\triangleq \mathbf{S}\mathbf{\Gamma}^{\frac{1}{2}}\mathbf{H} + \mathbf{W},\label{eq:preamble_signal}
\end{align}
where $\mathbf{S}=\left[\mathbf{s}_1,\ldots,\mathbf{s}_N\right]\in\mathbb{C}^{L\times N}$ is the matrix of fixed preamble sequences as columns, $\mathbf{\Gamma}=\mathrm{diag}\left(\gamma_1,\ldots,\gamma_N\right)\in\mathbb{R}^{N\times N}$ with sparse diagonal entries $\gamma_n = a_n g_n$, $\mathbf{H}=\left[\mathbf{h}_1,\ldots,\mathbf{h}_N\right]^T\in\mathbb{C}^{N\times M}$, and $\mathbf{W}\in\mathbb{C}^{L\times M}$ is additive white Gaussian noise with elements drawn independently from $\mathcal{CN}(0,\sigma_W^2)$. The goal is to recover the set of user activities $\{a_n\}_{n=1}^N$, possibly along with $\{\mathbf{h}_n\}_{n=1}^N$, based on $\mathbf{Y}$. Depending on the different scenarios, we may design sparse recovery algorithms that assume $\{g_n\}_{n=1}^N$ are deterministically known by the base station, or that they are random quantities with a known prior distribution, or that they are completely unknown. Note that we assume that the transmission takes place within a single coherence block. In the following, we outline the two algorithms for recovering $\{a_n\}_{n=1}^N$ that dominate the literature on massive random access, namely approximate message passing (AMP) and covariance-based decoding. The two algorithms differ in that the former provides an estimation of the channels in addition to recovering the user activities, but the latter can accommodate a much larger number of active users.

\subsubsection{Approximate Message Passing Decoding}
AMP~\cite{donoho2009message} is a low-complexity algorithm for sparse recovery in compressed sensing. Applied to the multi-antenna scenario in Eq.~\eqref{eq:preamble_signal}, a multiple measurement vector (MMV) variant of AMP (MMV-AMP)~\cite{chen18sparse} can be used to recover the row-sparse matrix $\mathbf{X}=\mathbf{\Gamma}^{\frac{1}{2}}\mathbf{H}\in\mathbb{C}^{N\times M}$. Since $\mathbf{X}$ uncovers both the set of active users and their channels, the MMV-AMP algorithm is particularly suitable for the grant-free scenario since no separate channel estimation phase is needed. Starting with $\mathbf{X}^0=\mathbf{0}$ and $\mathbf{Z}^0=\mathbf{Y}$, MMV-AMP operates iteratively according to the recursion
\begin{align}
  \mathbf{X}^{t+1}&=\eta\left(\mathbf{S}^*\mathbf{Z}^t+\mathbf{X}^t\right),\label{eq:mmvamp_s1}\\
  \mathbf{Z}^{t+1}&=\mathbf{Y}-\mathbf{S}\mathbf{X}^{t+1}
  + \frac{N}{L}\mathbf{Z}^t\left\langle\eta'\left(\mathbf{S}^*\mathbf{Z}^t+\mathbf{X}^t\right)\right\rangle,\label{eq:mmvamp_s2}
\end{align}
where $\eta(\cdot)$ is a vector denoiser with derivative $\eta'(\cdot)$ applied row-wise, and $\langle\cdot\rangle$ denotes the average across the rows of the argument. The first step in MMV-AMP (Eq.~\eqref{eq:mmvamp_s1}) can be interpreted as a gradient step followed by a denoising operation to promote sparsity in the estimate of $\mathbf{X}^{t+1}$. The second step (Eq.~\eqref{eq:mmvamp_s2}) computes the resulting residual corrected by the so-called Onsager term that transforms the statistics of the residual to enable faster convergence.

The choice of the denoiser function in Eq.~\eqref{eq:mmvamp_s1} depends on the specific model assumptions and the goal. Traditionally, the denoiser is implemented as a soft thresholding function, which is suitable when the activation probabilities and $\{g_n\}_{n=1}^N$ are treated as unknown quantities. However, if the statistics of these quantities are known to the decoder, it is possible to derive minimum mean-square error denoisers, which tend to have better overall performance~\cite{chen18sparse}. The complexity of MMV-AMP iterations is dominated by the computation of $\mathbf{S}\mathbf{X}^{t+1}$, and thus MMV-AMP has a complexity in the order of $O(NLM)$ per iteration~\cite{liu18part1}.

The preamble symbols are often chosen as independent and identically distributed zero-mean complex random variables with variance $1/L$. In this case, the evolution of the MMV-AMP algorithm can be accurately predicted through state evolution for the asymptotic regime where $L,N,K\to\infty$ while $N/L$ and $K/N$ remain constant~\cite{chen18sparse,liu18part1}. As a rule of thumb, the AMP algorithm works well in the regime where the number of users $K$ is less than the preamble length $L$. This is due to the fact that the AMP algorithm attempts to recover $K$ channels based on $L$ observations.

\subsubsection{Covariance-Based Decoding}
The covariance-based recovery method was proposed in~\cite{fengler21tit} as an alternative to MMV-AMP for the case where the total number of preambles is very large. Furthermore, it tends to perform better than AMP in many regimes, including when the number of antennas is large. Contrary to AMP, the covariance-based approach does not recover the fading coefficients $\{\mathbf{h}_n\}_{n=1}^N$ but only the user activities $\{a_n\}_{n=1}^N$. This makes the technique less suitable for grant-free random access, where channel estimates are required to decode the user messages. Finally, covariance-based decoding does not require knowledge of the large-scale fading statistics and the activation probabilities.

The covariance-based approach arises from the observation that the received signal at each antenna are independent and identically distributed realizations of a multivariate complex Gaussian distribution with zero mean and covariance matrix $\mathbf{\Sigma}=\sum_{k=1}^N \gamma_k\mathbf{s}_k\mathbf{s}_k^{H}+\sigma_Z^2\mathbf{I}$, where the superscript $H$ stands for conjugate transposition. The likelihood function of the received signal $\mathbf{Y}$ given $\boldsymbol{\gamma}=\left[\gamma_1,\ldots,\gamma_N\right]\in\mathbb{R}^{N}$ can then be computed as
\begin{align}
  p(\mathbf{Y}|\boldsymbol{\gamma}) &= \prod_{m=1}^M\frac{1}{|\pi\mathbf{\Sigma}|}\exp\left(-\mathbf{y}_m^H\mathbf{\Sigma}^{-1}\mathbf{y}_m\right)\\
  &=\frac{1}{|\pi\mathbf{\Sigma}|^M}\exp\left(-\tr\left(\mathbf{\Sigma}^{-1}\mathbf{Y}\mathbf{Y}^H\right)\right).
\end{align}
Although maximizing the likelihood leads to a non-convex problem, it is possible to obtain good solutions using a coordinate descent algorithm. The resulting set of active users can then be recovered from $\boldsymbol{\gamma}$. By analyzing the coordinate descent, it can be shown that the computational complexity is $O(TNL^2)$, where $T$ is the number of iterations~\cite{fengler21tit}. In practice, the covariance-based method tends to have longer runtime than AMP for small values of $M$, but is more efficient for larger $M$~\cite{chen22phase}.

Compared to AMP, performance analysis of the covariance-based approach is more involved, but have been characterized in~\cite{chen22phase} for the asymptotic case where the number of antennas goes to infinity while $N$, $K$, and $L$ remain finite. In this regime, it is possible to numerically obtain the detection error. This analysis reveals that the algorithm exhibits a phase transition, where the decoding error vanishes for a specific region of $N$, $K$, and $L$. As a key advantage as compared to AMP, the covariance-based approach can work well in the regime where the number of users $K$ is in the order of the \emph{square} of the preamble length $L$. This is due to the fact that the covariance matrix is of size $O(L^2)$. However, such advantage comes at a cost of not being able to estimate the channels. Finally, we remark that while the discussion in this section has been confined to the single-cell case, both the algorithmic developments and the analyses of AMP and the covariance method can be extended to the multi-cell or the cell-free massive multiple-input multiple-output (MIMO) scenarios~\cite{chen19multicell,chen21multicellfading,cakmak23cellfree,wang23cellfree}.

\subsection{Massive Downlink Connectivity}\label{sec:massivedownlink}
The majority of the research in massive connectivity has been put into the uplink scenario and the design of protocols for massive random access. However, the increasing amount of downlink traffic required by future IoT applications necessitates efficient communication schemes for the massive downlink scenario as well. In massive downlink communication, the transmitter, i.e., the base station, aims to deliver a set of messages to a small subset of a large user population. Because the set of recipients is unknown to the users a priori, the messages must include some form of metadata that allow each recipient to extract its own message. A naive implementation would be to include a list of the recipient identifiers, which would require a total of $K\log_2 N$ bits of metadata. For large values of $N$, this cost can be prohibitive, especially when the messages are short in comparison. However, by exploiting the structure of the problem, it is possible to reduce the amount of metadata significantly.

In a massive downlink channel, the base station aims to jointly send $N$ messages $W_1,\ldots,W_N$ to users $1,\ldots,N$, respectively, where $W_n\in\mathcal{M}\cup\{\varnothing\}$ for $n=1,\ldots,N$. Each message $W_n$ ($n=1,\ldots,N$) can either be \emph{informative}, drawn from the message set $\mathcal{M}$, or \emph{empty}, denoted by $W_n=\varnothing$, indicating that the transmitter has no message to user $n$ or that user $n$ is not actively listening to the base station. In line with the overall massive connectivity scenario, we assume that only a small random subset comprising $K$ out of the $N$ messages are informative, i.e., the fraction of empty messages ($W_n=\varnothing$) is equal to $\frac{N-K}{N}$. This formulation is reminiscent of the classical broadcast channel, which has been thoroughly studied in information theory. However, the regime that is of interest in massive downlink differs from the widely studied broadcast channel in the following ways:
\begin{itemize}
\item The number of users, $N$, is very large, often in the order of several thousands;
\item The set of informative messages is small (i.e., the messages are short), and only a small fraction of the messages are informative, leading to a highly skewed message distribution.
\end{itemize}

In the rest of this section, we consider three instances of the massive downlink problem, and demonstrate that, in these fairly general cases, it is possible to significantly reduce the amount of metadata by jointly encoding the messages into a single packet. Specifically, the dependency on $N$ can be reduced from $O(\log N)$ to $O(\log\log N)$ or even $O(1)$ for fairly general problem setups. To keep the presentation clear, we focus on the source coding aspect of the problem and assume that the channel is ideal, i.e., $Y_n=X$ for $n=1,\ldots,N$. However, we note that joint encoding is also beneficial from a channel coding perspective, as the larger packet size allows for more efficient coding.

\subsubsection{Message Acknowledgments}\label{sec:messageacks}
Suppose the base station wants to send binary message acknowledgements to $K$ out of the $N$ users, e.g., after an uplink massive random access transmission. This corresponds to the case where $|\mathcal{M}|=1$, so that the base station sends the empty message as a negative acknowledgment and the informative message as a positive acknowledgment. When $K$ is much smaller than $N$, it might seem reasonable to construct the acknowledgment packet by concatenating the identifiers of the users to acknowledge, which would require $K\log_2 N$ bits. However, as shown in~\cite{kalor22tcom}, it is possible to significantly better by enumerating all possible subsets of users to acknowledge, which requires only $\log_2\binom{N}{K}\le K\log_2(N/K)+K\log_2 e+O(1)$ bits. Nevertheless, the number of bits can still be significant when $N$ is large. For instance, acknowledging $K=50$ users, each identified by a $64$-bit identifier ($N=2^{64}$), still requires more than $2900$ bits.

Reducing the packet length further requires that errors are allowed. When $K$ is much smaller than $N$, allowing for a small fraction of false positive acknowledgments can lead to significant savings, whereas false negative errors provide limited gains. While false positive acknowledgments may be undesirable from an application perspective, they can be corrected at higher layers in the protocol stack at the cost of an increased latency using, e.g., packet sequence numbers. By allowing false positive errors with probability at most $\epsilon$, it is possible to construct an acknowledgment packet using as little as $K\log_2(1/\epsilon)+O(\log\log N)$ bits~\cite{kalor22tcom}. Remarkably, the introduction of false positive errors makes the acknowledgment message almost independent of $N$. Furthermore, this can be realized using practical methods that are efficient up to relatively large values of $K$ (in the order of hundreds). Disregarding the constant term (which is very small in practice), acknowledging $K=50$ users with $\epsilon=10^{-4}$ requires a packet of only $665$ bits, which is significantly less than in the error-free case.

\subsubsection{Transmission of General Feedback Messages}
We consider next the problem of transmitting $K$ general messages $W_{1'},W_{2'},\ldots,W_{K'}\in\mathcal{M}$ to $K$ out of the $N$ users, which was recently studied in~\cite{song23isit,song24massivedl}. Compared to the previous case, instead of assuming a false positive rate, we will assume that the devices know whether their own message is informative or empty, but have no information about the other messages. This assumption is often reasonable, e.g., if the messages are transmitted as responses to an uplink transmission. Alternatively, the message acknowledgment scheme presented in Section~\ref{sec:messageacks} can be used to indicate the intended recipients. The central problem is then to encode the $K$ messages in such a way that the recipients know how to decode their own message without knowing the identities of the $K-1$ other recipients of informative messages. Note that, if the users know perfectly the identifiers of all $K$ recipients, this can be done simply by listing the messages in the order of the identifiers. However, when each user only knows whether there is a packet for them or not, and not who the other recipients are, the problem becomes more involved. Assuming that the messages are independent and identically distributed across users, the naive encoding scheme of concatenating each of the $K$ messages with the identity of its recipient requires approximately $kH(W_{n})+K\log_2 N$ bits, where $H(W_n)$ denotes the entropy of a message $W_n$. However as shown in~\cite{song23isit}, theoretically speaking, the overhead of $K\log_2 N$ bits can in fact be reduced to only $O(\log_2 K)$ bits for any independent and identically distributed (i.i.d.) messages, and further to $O(1)$ when the messages are equally likely across the alphabet, thus removing the dependency on $N$.

\subsubsection{Collision-Free Scheduling}
Finally, we consider the problem of communicating a specific type of a collision-free scheduling message that assigns $K$ out of $N$ users to $B\ge K$ indices without collisions. This problem is central in, e.g., grant-based random access, where the base station needs to assign uplink resources to the users after the initial preamble transmission. A naive implementation would be to transmit an ordered list of the $K$ identities and assign each user to the index at which their identity appears in the list. This would require a total of $K\log_2 N$ bits. However, as in the previous cases it is possible to improve significantly by jointly encoding the assignments and exploiting the fact that the users do not care about the indices assigned to other users. Furthermore, the scheduling problem differs from the previous two problems in that the ordering of the scheduled users does not matter as long as the schedule is non-colliding. By taking advantage of these characteristics, it was shown in~\cite{kang21tit} the packet length can be reduced to $K\log_2 e+O(\log\log N)$ bits for fixed-length encoding and $K\log_2 e+O(1)$ for variable-length encoding. As expected, the packet length can be further reduced when either $B>K$ or $B<K$, where in the latter case up to $\lceil K/B\rceil$ users are allowed in each slot. While no practical encoding method is known to achieve the theoretical bounds, techniques developed to construct minimum perfect hash functions can be used to achieve rates that are within a relatively small constant factor of the bounds~\cite{kang21tit}.

To summarize, communicating a downlink message in the massive random access scenario can theoretically be made much more efficient than what a straightforward implementation would suggest. These theoretical bounds point to significant potential for future protocol-level improvements for massive downlink random access in 6G and beyond.

\subsection{Integration of Massive Connectivity in 6G Standardization}\label{sec:standarziation}
So far, we have discussed massive connectivity technologies from a theoretical perspective. In this section, we discuss how these ideas can be integrated in 6G standardization, as well as some of the remaining challenges that need to be addressed.

Let us first focus on the integration of the enhanced technologies for random access schemes in the uplink presented in Section~\ref{sec:grantbased_grantfree}. A first step towards improving the performance would be to reuse the existing access structure in 5G, and augmenting it with, e.g., non-orthogonal preambles. This idea has recently been analyzed for grant-free access in~\cite{agostini24ra6g}, where the authors demonstrate that the inclusion of non-orthogonal preambles in the 5G SDT procedure substantially increases the energy and spectral efficiency, albeit still being far from the theoretical limits. In this case, the sparse recovery algorithms discussed in Section~\ref{sec:massive_cs} will play a central role in the decoding of the preambles. Furthermore, since grant-based random access also benefits from non-orthogonal pilots, this approach can also be adopted to improve the grant-based random access procedures.

In the context of downlink, the techniques discussed in Section~\ref{sec:massivedownlink} are more straightforward to realize, as they do not require modifications to the physical layer. For the same reason, they do not rely on specific assumptions that need to be validated experimentally. However, while it may be unlikely that downlink designs tailored for specific applications will be implemented in early 6G, the theoretical insights from Section~\ref{sec:massivedownlink} provide useful guidelines in the design of the control information for massive access in 6G, such as the design of acknowledgments and scheduling for grant-based access.

\section{Critical Services}\label{sec:critical}

Critical services in wireless communication have undergone a significant evolution, transitioning from 5G focus on mission-critical applications with stringent  reliability and latency requirements to the more advanced landscape of safety critical control applications. The emerging demands include real-time responsiveness, task-oriented communications, high availability, and resilience.

The spectrum of applications within this safety critical control domain is extensive, encompassing  autonomous vehicle platoons, advanced industrial automation and human augmentation. Fulfilling the stringent requirements of these applications necessitates integrating cutting-edge technologies, {such as appropriate spectrum, systems with massive number of antennas, and ML. The critical requirements may also benefit from some of the emerging technologies, such as reconfigurable intelligent surfaces (RIS) or the THz band.} 
Moreover, the sophisticated interaction between control and communication systems must be carefully considered during the design phase. The resultant complexity of this multifaceted problem underscores the need for the usage of AI. AI enables the optimization of resources and design in an adaptive manner, learning from near-real-time physical operating scenarios. 

However, the deployment of AI requires careful consideration. There is a need for the development of novel techniques aimed at enhancing the convergence time of the algorithms, especially in the context of improving network intelligence within strict delay constraints. Moreover, ensuring reliability and robustness while fostering continuous learning poses a challenge. In addition, fortifying the resilience of developed algorithms is imperative for sustained performance in dynamic environments.  

\begin{figure}
  \centering
  \includegraphics{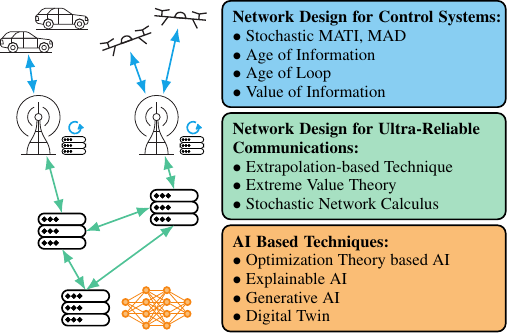}
  \caption{Enabling tools and technologies for supporting critical services in 6G.}\label{fig:critical_netw_design}
\end{figure}

In the following, we begin by describing the applications and requirements of critical services. Following this, we delve into the enabling technologies essential for supporting these critical services. Subsequently, we provide the network design to meet the specific requirements of control applications within the critical services domain. Finally, we illustrate the usage of AI based techniques, emphasizing their role in generating solution methodologies that effectively address the stringent requirements of delay, reliability, and resilience inherent to these applications. An overview of the key enablers is given in Fig.~\ref{fig:critical_netw_design}.

\subsection{Applications for Critical Services}

The applications of critical services in 6G introduce heightened demands for superior delay, reliability, availability, and resilience. These applications closely interact with control systems, necessitating real-time responsiveness and task-oriented communications. As the realm of 6G progresses, the potential applications for critical services continue to evolve. A few representative examples are listed as follows:

\begin{itemize}\setlength\itemsep{1mm}
\item \emph{Autonomous vehicle platoon:} An autonomous vehicle platoon refers to a convoy of self-driving vehicles seamlessly communicating in real-time for synchronized acceleration, braking and steering. The communication platform supporting these platoons necessitates careful consideration of their close interaction with automotive control systems. Real-time responsiveness is paramount, demanding an exceptionally high level of reliability, availability, and resilience, along with guaranteed low delay. This robust communication framework is essential to prevent any disruptions in platoon operations, ultimately ensuring the utmost safety for the occupants of the vehicles.    
\item \emph{Advanced industrial automation:} Industrial automation {was targeted by 5G through multiple applications, such as} remote monitoring, predictive maintenance, and smart manufacturing. 6G elevates this paradigm to new heights by introducing applications with even more stringent demands. Specifically, it enables real-time control of robotic systems, facilitates collaboration among robots and orchestrates the seamless integration of humans and robots in safety-critical processes. This evolutionary step in 6G requires a close examination of the interaction between communication systems, and industrial and robotic control systems. It aims to provide exceptional performance in terms of delay, reliability, and availability, leading to a new era of precision and efficiency in industrial automation. 

\item \emph{Human augmentation:} Building upon the foundation laid by 5G, 6G represents a significant leap forward in the realm of sensory augmentation devices, including smart prostethics, implants and wearables.  Going beyond the capabilities of its predecessor, 6G advances sensory augmentation through its close interaction with precise and real-time control, showcasing significantly improved latency, reliability, and availability performance. This evolution broadens the spectrum of applications, enabling the precise control of augmented limbs, the delivery of more realistic haptic feedback and the attainment of higher resolution bionic vision and hearing.

\end{itemize}

\subsection{Enabling Technologies}

Critical services in 6G necessitate the adoption of cutting-edge technologies to meet the stringent demands for reliability, low latency, availability, and resilience. 
In general, critical communication requires sources of diversity in all forms, including spectrum, redundant equipment, failover mechanisms, and, not the least, large number of antennas. Additional diversity can be sourced from some of the emerging technologies, such as RIS with large number of reflective elements, as well as integration of spectrum with very high frequencies, such as THz and visible light communication. Some of these technologies are discussed in the sequel.

Distributed MIMO and cell-free massive MIMO represent {an evolution of the} traditional cellular network architecture, transforming into a collaborative system of a multitute of access points (APs), each equipped with multiple antennas per AP and distributed across the coverage area. This {type of} deployment significantly improves the reliability and delay performance of communication links by minimizing the average distance between users and APs due to the spatial distribution of multiple antennas \cite{cellfreemimo_urllc1}. Moreover, this cell-free architecture accommodates diverse reliability, delay and availability requirements of multiple users simultaneously through spatial multiplexing over which multiple users can be concurrently served using the same frequency resources \cite{cellfreemimo_urllc2}. Transmit beamforming over multiple spatially distributed antennas  is another key feature,  enabling the system to precisely direct signals {towards the} intended receivers while concurrently minimizing interference for other users \cite{cellfreemimo_urllc3,cellfreemimo_urllc4}. Finally, there is also the possibility of exploiting diversity at the network level by taking advantage of multi-hop cooperative relaying opportunities, which can be effective especially in scenarios with blockage~\cite{khosravirad21diversity}.

At this stage, the use of RIS and THz frequencies in 6G is more speculative, although there are prototypes and demonstrations of benefits of these technologies in specific setups. RIS leverage programmable materials, with configurable phase and amplitude, to alter the reflection behavior of the incident waves. This dynamic control of the signal propagation serves to mitigate fading and interference, thus fostering heightened reliability in communication while ensuring exceptional availability \cite{ris_urllc1, ris_urllc2}. In the context of critical services, RIS can also optimize signal propagation paths and facilitate of precise beamforming techniques \cite{ris_urllc3}.

Finally, THz
communications encompass the transmission of signals within the frequency range of $100$ GHz to $10$ THz, offering a significantly increased bandwidth compared to lower frequency bands. This ultra-high bandwidth capacity facilitates the transmission of data at exceptionally high data rates, paving the way for applications that demand both high rate and high reliability, such as wireless virtual reality \cite{THz_urllc1}. However, THz communication faces a significant challenge: high atmospheric absorption by atmospheric gases, leading to signal attenuation. Nevertheless, recent advances have addressed this challenge by developing techniques to select frequency bands with lower absorption rates \cite{THz_frequencyselection}, utilize beamforming to avoid regions of high absorption \cite{THz_beamforming}, and develop adaptive modulation and coding schemes to minimize errors induced by absorption \cite{THz_adaptivecoding}. These advances have been demonstrated to achieve multi-kilometer transmission ranges \cite{THz_km}. Although THz communications do have a potential to be combined with RIS, its adoption in 6G standards may still be premature, leaving its use open for future wireless generations.

\subsection{Network Design for Critical Services in 6G}

The evolution of critical services, transitioning from mission-critical applications with stringent reliability and latency requirements to safety critical control applications, necessitates an enhanced approach to network design. This progression calls for the integration of advanced control system-aware performance metrics, ensuring not only high availability and resilience but also a shift towards task-oriented communications. 
{Moving beyond the usual guarantees for correct reception of }individual transmitted bits, the network design for critical services should consider the impact of each {data portion} on the stability and performance of the associated control systems. Furthermore, meeting the requirements of high availability and resilience entails a comprehensive grasp of the statistical properties of the channel, queue length, and delay. By modeling the tail statistics of these parameters, the network design can better account for extreme events and fluctuations, leading to a more robust and adaptive system architecture.

\subsubsection{Network Design for Control Systems}

The joint design of control and communication systems requires a thorough examination of their dynamic interdependencies. Key considerations include the sampling pattern within the control system, along with the patterns of message delay and dropout within the communication system. Achieving seamless integration of these two systems requires a detailed exploration of control system abstractions, as well as {working with more general timing measures, rather than focusing only on latency~\cite{popovski22timing}.} The following details various such abstractions {and some timing measures}:

\begin{itemize}\setlength\itemsep{1mm}

\item \emph{Stochastic maximum allowable transfer interval (MATI) and maximum allowable delay (MAD):} The criteria governing the performance and stability of control systems are expressed through the definition of two key parameters: MATI and MAD \cite{matimad_1, matimad_2}. While wireline networks can often meet these rigorous real-time criteria, the same expectations are infeasible in wireless networks due to inherent packet error probability. Consequently, wireless control systems adopt a stochastic MATI constraint, ensuring that the time interval between successive state vector reports remains above MATI value with a predefined probability \cite{stochasticmati_1, stochasticmati_2}. 

The incorporation of stochastic MATI and MAD constraints have been recently explored in the context of joint design with communication systems \cite{stochasticmati_2, stochasticmati_3, stochasticmati_4}. In \cite{stochasticmati_2}, these constraints are incorporated into the optimization problem aiming to minimize power consumption while ensuring schedulability in the communication system. This initial formulation is tailored for M-ary Quadrature Amplitude
Modulation (MQAM) as the modulation scheme and Earliest
Deadline First (EDF) as the scheduling algorithm. Building on this foundation, \cite{stochasticmati_3} extends the framework to accommodate any modulation scheme and scheduling algorithm. Both \cite{stochasticmati_2} and \cite{stochasticmati_3} propose efficient polynomial time heuristic algorithms, leveraging optimality conditions, relaxation and search space reduction techniques to achieve close-to-optimal performance. The demonstrated success of the proposed joint design methodology highlights its superiority over the traditional practice of separately designing control and communication systems. This holds true across a diverse range of network and control system parameters, affirming its robustness and adaptability in varying conditions.

Taking the exploration further, \cite{stochasticmati_4} incorporates ultra-reliable transmission in the finite blocklength regime. The proposed solution methodology adopts an optimization theory based Deep Reinforcement Learning (DRL) approach, comprising two stages: an
optimization theory stage where optimality
conditions are derived to reduce the decision
variables, and a DRL stage where DRL is employed to tackle specific parts of the optimization problem that are not tractable. This AI based approach ensures the adaptivity of the solution in stochastic environments. 

\item \emph{Age of Information (AoI):} AoI is a metric that measures the time elapsed since the latest relevant information was generated for a state update at the controller. 
The effectiveness and stability of the control processes hinge on adhering to constraints on AoI statistics, ensuring that the control system operates with the latest and most pertinent state information. The analysis and optimization of AoI statistics has attracted significant attention across diverse communication network scenarios, as evidenced by studies such as~\cite{aoi_1}. 

Recent research has delved into connecting the AoI of sensor data with the control system performance \cite{aoi_2, aoi_3, aoi_4}. Specifically, \cite{aoi_2} proposes a deep learning (DL)-based framework for co-designing estimators, controllers and schedulers, incorporating awareness of the sensor's AoI and dynamic channel states. This framework introduces an AoI-based importance sampling algorithm, enhancing learning efficiency by considering both data accuracy and freshness. Targeting stable mean-square estimation (MSE) performance over short periods for time-critical systems, \cite{aoi_3} limits AoI growth through optimizing Hybrid Automatic Repeat Request (HARQ) packet retransmission schemes. This involves allowing partial retransmissions and optimizing retransmission times to facilitate the early arrival of subsequent status updates. Additionally, \cite{aoi_4} studies optimal scheduling to minimize overall estimation MSE, deriving structural properties of the optimal sensor scheduling policy based on AoI states and corresponding channel states. This policy is integrated into a deep reinforcement learning (DRL) framework, enhancing effective exploration of the action space.

\item \emph{Age of Loop (AoL):} AoL is defined as the elapsed time since the generation of information leading to the latest action or state update in a wireless networked control system (WNCS). Unlike AoI, which is defined for a single communication direction, either downlink or uplink, AoL is considered
a more relevant metric for optimizing closed-loop
WNCS by considering both uplink and downlink of the control loop \cite{aol_1, aol_2}. In \cite{aol_1}, the AoL metric is employed to learn the WNCS latency and freshness bounds. The study proposes an AoL-based bandwidth allocation policy aimed at minimizing the WNCS cost. Results show that the proposed algorithm outperforms policies based on fixed latency requirements. Expanding on this, \cite{aol_2} conducts a thorough investigation of the peak age of loop (PAoL) performances. This includes the average, variance and outage probability of PAoL for WNCSs operating in the finite blocklength regime over fading channels. This comprehensive analysis paves the way for a PAoL-oriented communication and control co-design, introducing an adaptive scheme for transmission power, blocklength and the maximum number of allowable transmissions. The proposed PAoL-oriented scheme demonstrates significant performance gains when compared to UL-only and DL-only optimization schemes.

\item \emph{Value of Information (VoI):} VoI serves as a crucial metric, capturing the variation in a value function concerning the information available to the controller about the state of a process. Distinguished from AoI and AoL, which focus on the timeliness of the updates, VoI delves into the content of the information, measuring the semantics of each data packet \cite{voi_1, voi_2, voi_3}. In \cite{voi_1}, VoI is quantified as the difference between the benefit derived from enhancing regulation quality due to reduced uncertainty at the controller and the transmission cost of a packet. This study introduces a strategy based on VoI, demonstrating optimal management of the communication between the sensor and controller. It is worth noting that this research assumes negligible network-induced effects, such as transmission delay, quantization or packet dropouts. Extending this concept to networked control systems with arbitrary transmission delay, \cite{voi_2} defines VoI metric as the reduction of uncertainty from the decision maker's perspective given a measurement update. This VoI metric serves as the triggering condition for NCS closed over a communication network. The proposed VoI based event-triggering policy proves to achieve a lower MSE compared to the AoI-based and periodic updating policies. In the context of an event-triggered control system with multiple subsystems sharing a wireless channel, \cite{voi_3} proposes a channel access prioritization scheme based on
the VoI. The VoI of a subsystem depends on channel reliability and
represents the difference between sensor’s a posteriori state
estimate and the estimator’s estimate in the absence of any data. The objective of the proposed scheme is to minimize the discrepancy in the state estimate. Utilizing measurement based triggering and VoI as the priority measure leads to optimal performance. 

 \end{itemize}

 The exploration of the joint design of control and communication systems is currently in its early stages. A comprehensive comparison of control system abstractions across various application scenarios remains an open research topic. While the consideration of all the system parameters and realistic assumptions in the system model is crucial for optimal network design, it has been largely neglected in most designs due to increasing complexity. Dealing with the resulting high complexity of these problems necessitates the application of adaptive machine learning techniques. 

\subsubsection{Network Design for Ultra-Reliable Communication}

Critical services often demand URLLC with a targeted packet error rate ranging from $10^{-9}$ to $10^{-5}$, and acceptable latency on the order of milliseconds or less. 
{In certain cases, depending on the fading and knowledge of channel state information, the Shannon capacity formula applicable to infinite blocklengths may not be sufficiently accurate when finite blocklength is considered~\cite{durisi2016toward}.} In \cite{fbl_1}, rigorous bounds as well as an approximation for the rates achievable in the finite-blocklength regime has been derived, considering signal-to-noise ratio (SNR), blocklength, and decoding error probability. This expression has further been extended to fading channels in \cite{fbl_1, yang14-07c}. Recent studies on URLLC have integrated these finite blocklength rate expressions into their designs, focusing on methodologies to extend the original approximation to practically relevant transceiver architectures and  to evaluate such an approximation efficiently~\cite{fblurllc_1, fblurllc_2, fblurllc_3, lancho20-07j,lancho23-12a, kislal23-07a}.

{Ensuring ultra-high reliability involves tackling fundamental challenges in the statistical modeling of the wireless channel, queue length, and delay within the ultra-reliable region.} Meeting the stringent reliability requirement of URLLC  necessitates innovative techniques to analyze the lower tails of distributions, especially when dealing with extremely low probabilities. Additionally, efficiently handling and optimizing a large amount of data are crucial for modeling infrequently occurring extreme events. Deriving tail statistics might be perceived as collecting a large amount of data and fitting it to probabilistic distributions. However, these distributions may fall short in capturing rare events due
to limited data or may not be applicable in different environmental conditions. Furthermore, since diversity techniques (time, frequency, interface, and spatial diversity) are employed to reduce the required signal-to-noise-ratio (SNR) for achieving a specific reliability, tail statistics must account for the dependencies of extreme events in multiple dimensions. Various statistical methods have been employed to address these challenges, as summarized below: 

  \begin{itemize}   
        \item \emph{Extrapolation based technique:}
The extrapolation based technique relies on establishing  a unified statistical representation for a diverse set of wireless channels operating in the ultra-reliable regime of operation \cite{extrapolation_1}. This method  approximates the behavior of wireless channels in the ultra-reliable regime using simple power law expression, where the exponent and offset are contingent on the specific channel model. The framework is extended to accommodate multiple receivers by incorporating the  simplified expression into maximum ratio combining (MRC).

Building upon this foundation, \cite{extrapolation_2} further extends the study by introducing a rate selection framework tailored for URLLC systems. This framework comprises three essential components: channel model selection, model learning through training; and transmission rate selection to meet
the required reliability. Recognizing the inadequacy of specifying URLLC requirements solely through packet error rates, two types of statistical constraints are introduced: averaged reliability, governing the mean outage probability across all possible realizations of the training sample, suitable for designing
URLLC systems with desired average performance, and probably correct reliability, which manages the probability that the
outage probability surpasses a specified threshold for a given training sample. These extrapolation-based approaches prove effective when the extrapolation of average-statistics channel models aligns with empirical data in the ultra-reliable region. 

Collecting a sufficient number of samples from a device at a given spatial point may require excessive number of channel samples. Extrapolation-based approaches can be extended to leverage the past data collected at nearby spatial points in order to make reliability prediction for a given spatial point. This is the main idea behind statistical radio maps~\cite{kallehauge2023delivering}, where reliability prediction relies on assumptions for spatial smoothness.
        
\item \emph{Extreme value theory:} Extreme value
theory (EVT) stands out as a distinctive statistical discipline designed to devise methodologies and models for characterizing rare events in ultra-reliable communication through the application of mathematical limits as finite-level approximation. EVT finds applications at physical layer to model the tail statistics of the channel and design ultra-reliable communication systems based on these tail statistics \cite{evt_1, evt_2, evt_3, evtrate_1, evtrate_2}. Additionally, it is employed at data link and network layers to model the tail statistics of queue length and delay, and incorporate these tail statistics into the optimization of resource allocation \cite{evtdelay_1, evtdelay_2, evtdelay_3}.

In \cite{evt_1}, EVT techniques are used at the physical layer to determine the optimum threshold below which the received power samples are classified as extreme events and incorporated into the tail distribution. The generalized Pareto distribution (GPD) is then employed to model this tail distribution. The validity of the proposed EVT-based model is assessed through the examination of probability plots. Building upon this, \cite{evt_2} extends the methodology to characterize the tail of the non-stationary channel distribution. This is achieved by identifying external factors contributing to non-stationarity and segmenting the channel data into multiple stationary sequences. The parameter
of the fitting distribution is modeled as a change-point function with respect to time. Furthermore, \cite{evt_3} takes these techniques a step further for systems employing spatial diversity. It calculates the lower tail statistics of received signal power across multiple dimensions, efficiently managing extensive corresponding data. The approach involves fitting bi-variate GPD to the tail of the joint probability distribution, employing logistic distribution-based and Poisson point process-based approaches. The validity of these approaches is evaluated by incorporating the probability measure function of the Pickands coordinates.

Moving to real-time ultra-reliable communication, \cite{evtrate_1} proposes a framework for rate selection using EVT. The maximum transmission rate is determined by integrating GPD into the rate selection function. This framework is expanded in \cite{evtrate_2} by introducing confidence intervals to the estimations, addressing cases with limited data availability. The outcomes of these studies indicate that the proposed rate selection framework provides a feasible approach to reach a specified target error probability. This is accomplished by using a higher transmission rate and reducing the amount of training data, as opposed to conventional rate selection methods.

Transitioning to data link and network layers, \cite{evtdelay_1}, \cite{evtdelay_2} and
\cite{evtdelay_3} employ EVT principles to analyze the statistical properties of large queue lengths
and AoI, and integrate them into resource allocation problem. In \cite{evtdelay_1}, the
distribution of the maximal queue length across a vehicular network is characterized through the application of EVT.
These findings are integrated into a stochastic optimization problem aimed at minimizing total
power. This methodology effectively contributes to lowering both the mean and variance of the
maximal queue length within the network. On the other hand, \cite{evtdelay_2} focuses on
characterizing the tail distribution of AoI using EVT. This is subsequently utilized to formulate a
transmit power minimization problem, considering probabilistic AoI constraints. The study sheds
light on the tradeoff between the arrival rate of status updates, and the average and worst AoI
achieved by the network. Characterizing the distribution of extreme events using EVT in the above
works necessitates the acquisition of sufficient samples capturing extreme events, which may
introduce unacceptable overheads in the process of gathering samples over the network. To address
this problem, \cite{evtdelay_3} proposes a distributed approach based on federated learning,  where the vehicular users estimate the tail distribution of the network-wide queues locally without
sharing the actual queue length samples. The proposed method has been demonstrated to learn the statistics
of the network-wide queues with high accuracy.

\item \emph{Stochastic network calculus: }  Stochastic network calculus (SNC) 
  offers robust theoretical insights into the analysis of tail
  distributions for ultra-reliable communications. Within this framework, non-asymptotic statistical
  performance bounds on delay, AoI, and reliability are computed, while considering complex
  stochastic processes \cite{snetworkcalculus_1, snetworkcalculus_2, snetworkcalculus_3}. In
  \cite{snetworkcalculus_1}, a comprehensive procedure is presented for calculating the end-to-end
  delay violation probability for target traffic in industrial Internet of Things by using SNC and
  moment generating functions. The analysis explores the impacts of various resource allocation
  strategies and parameters on probabilistic end-to-end delay. An
  SNC-based model to derive an upper bound for packet transmission delay, given the amount of
  allocated radio resources for a URLLC slice, the target violation probability and the distribution
  of traffic demand is proposed in \cite{snetworkcalculus_2}. The study proposes heuristics for the strategic planning of URLLC slices. It
  determines the optimal allocation of radio resources to each slice, ensuring that its delay bound
  aligns with the specified delay budget. Furthermore, \cite{snetworkcalculus_3} introduces a
  non-orthogonal multiple access-assisted uplink URLLC network architecture, incorporating a
  SNC-based statistical quality of service-provisioning theoretical framework to facilitate tail
  distribution analysis concerning delay, AoI, and reliability. The study formulates a power
  optimization problem aimed at minimizing transmit power while satisfying ultra-reliability
  constraints on delay, AoI, reliability, and power consumption.

\end{itemize}

The investigation of tail statistics for ultra-reliable communication systems is currently in its initial phases. A thorough examination of system behavior, taking into account tail statistics across multiple layers, remains an ongoing research area. Moreover, the integration of the tail statistics and finite blocklength information theory into the network design for control systems has been primarily explored in the context of AoI. To establish a more comprehensive understanding, further research is needed to connect these studies with control system performance. Furthermore, efficient derivation of these tail statistics with minimal overhead calls for the application of novel machine learning techniques, including transfer learning and federated learning. 
As a final remark, the presented tools and methods should be used as a basis to design actual procedures for assessing ultra-reliability in communication protocols. Eventually, all reliability guarantees should be data-driven, where the data is collected in a way that can capture the specifics of the communication setup and can deal with the changes in the data statistics.

\subsection{AI Based Techniques for Critical Services}

The growing demands of critical service applications, characterized by stringent requirements in terms of delay, reliability, availability and resilience, coupled with the close interaction of these applications with control systems, and the ongoing trend towards deploying massive MIMO and RIS systems at higher frequency bands, pose substantial challenges in the effective management of network functions and wireless resources. The resulting complexity in network design surpasses the capabilities of conventional analytical methods for modeling and optimization. Therefore, leveraging data-driven AI approaches becomes imperative for the network design of critical services. These AI-based approaches are instrumental in addressing the complexity and potential intractability of problem formulations, while simultaneously enhancing adaptability through the learning of near-real-time communication characteristics. 

Despite the remarkable performance of AI, the application of AI techniques in critical services
necessitates considerable caution. AI based approaches demand extensive data for training, giving
rise to significant communication delay. Moreover, the resulting models often rely on intricate deep
learning (DL)-based models and algorithms, characterized by their black-box nature, making them
challenging to understand and interpret. This lack of interpretability raises concerns about their
suitability across diverse scenarios. Potential performance issues may arise when confronted with
new data that deviates from the training set, thereby affecting resilience. Additionally, the
exploration of various actions in these techniques may have safety implications, introducing risks
to the reliability and robustness of system. The deficiencies in delay, interpretability,
robustness, reliability, and resilience of AI-based techniques present a significant risk in their deployment for critical services. To address these concerns, various techniques have been proposed, as detailed below:

\begin{itemize}\setlength\itemsep{1mm}

\item \emph{Optimization theory based AI:} Optimization theory based AI approach aims to leverage optimization theory knowledge to improve the performance of AI algorithms. This reduces the reliance on extensive training data, leading to minimized delays and improved solution reliability and robustness \cite{optdl_1, optdl_2, stochasticmati_4, optdl_4}. In \cite{optdl_1}, a deep neural network (DNN) is first pretrained using synthetic data generated from an optimization-theory based algorithm, followed by fine-tuning in a real environment with actual measured data. In DNN, system parameters are mapped to optimal resource allocation. Pretraining DNN involves generating a diverse set of system configurations, representing all possible inputs, and computing the optimal resource allocation using the optimization-theory based algorithm. This pretraining substantially decreases the amount of real-time data required for the implementation of AI approaches.

In \cite{optdl_2}, \cite{stochasticmati_4} and \cite{optdl_4}, the optimization theory based model breaks down algorithms into fundamental building blocks, replacing them with DNN architectures rather than treating the entire algorithm as a black box. Specifically, \cite{optdl_2} introduces a low-complexity
optimization theory-based deep learning framework for minimizing the schedule length in net-zero-energy networks with short packets. By deriving optimality conditions, mathematical relations among optimal decision variable values are established. Consequently, only a subset of outputs requires direct DNN computation, reducing training time by simplifying the complexity of the DNNs. On the other hand, \cite{stochasticmati_4} proposes an optimization theory based DRL framework for the joint design of controller and communication systems. Optimality conditions illustrate the mathematical
relations between optimal decision variable values, enabling the problem to be decomposed into multiple building blocks. Blocks that are simplified but not tractable are then replaced by DRL. The frameworks in \cite{optdl_2} and \cite{stochasticmati_4} are further generalized in \cite{optdl_4} by constructing the block diagram of any optimization theory-based solution, leveraging optimality conditions at both the input and output, and structure of iterative solutions from the realm of optimization theory. The proposed approaches not only reduce training time but also show significant enhancements in both overall accuracy and robustness performance in comparison to earlier propositions in optimization theory and deep learning.

\item \emph{Explainable AI:} Explainable AI (XAI) comprises processes and techniques designed to unveil the internal workings of AI algorithms to improve their interpretability, robustness and convergence time performance \cite{xai_1, xai_2}. XAI facilitates the creation of interpretable models by developing a locally reliable representation of the original black-box model behaviour using Local Interpretable Model-agnostic Explanations (LIME) method. Moreover, XAI plays a crucial role in evaluating and enhancing system robustness against adversarial attacks through sensitivity analysis and adversarial training. Sensitivity analysis quantifies the model output’s sensitivity to changes in input data, whereas adversarial training involves augmenting the training data with adversarial examples, improving decision reliability and resistance of the model to future attacks. In the context of DL-based robust beam alignment for MIMO networks, \cite{xai_1} demonstrates the effectiveness of XAI in improving robustness, with up to a 17-fold increase in the number of detected adversarial beam indices. XAI also contributes to reducing model complexity and shortening convergence time by providing techniques to eliminate less important input features and employing model compression techniques. The study in \cite{xai_1} showcases the application of XAI to decrease the average training time by $49.1\%$, while maintaining the performance gap less than $1\%$ for the DRL based resource allocation problem in vehicular networks. 

\item \emph{Generative AI:} Generative AI pertains to artificial intelligence systems designed to generate new and realistic data instances resembling and distinguishable from the training data. In the context of ultra-reliable communication for critical services, its primary application involves data augmentation during the pretraining of DRL algorithms using generative adversarial networks (GANs) \cite{generative_ai_1, generative_ai_2}. This aids in acquiring experience and learning extreme events, enhancing reliability, and attaining resilience. Utilizing GANs, a limited set of real traffic and wireless channel data is transformed into an extensive and diverse dataset, facilitating the training of DRL agents by providing experiential knowledge. The GAN-generated dataset effectively eliminates transient training times, enabling the agent to quickly adapt and recover during extreme conditions.

\item \emph{Digital twin:} Digital twin (DT) represents a digital replica of a physical entity, mapping real-world objects and environments to a virtual space in real time \cite{dt_1, dt_1_2}. Typically situated at the network edge to accommodate the low-latency requirements of critical services, DT undergoes an initial pretraining phase to optimize resources efficiently by employing mathematical optimization and machine learning techniques alongside virtual representation. The pretraining phase allows for exploration within a simulation platform, avoiding potential risky decisions, thus, enhancing the communication reliability. Subsequently, the DT leverages real-time data from the physical environment to enhance the training of the previously trained model through the application of transfer learning techniques. This addresses the challenges of model mismatch and nonstationarity in wireless networks. The pretraining significantly improves the convergence of the algorithms, thereby minimizing delays and enhancing the robustness of communication.

\end{itemize}
  
The ongoing exploration of AI-based techniques to fulfill the delay, robustness, reliability and resilience requirements of critical services remains an active and open area of research. Specifically, AI based techniques can only be expected to work as well as what the algorithms have seen in the training dataset, so it would be challenging for an AI based solution to be truly robust in the sense of being able to anticipate and to deal with \emph{unforeseen} rare events. Further, while innovative AI techniques have been introduced in specific contexts, there is a need for the design of new network architectures to establish a systematic methodology for their development and implementation. Moreover, the transferability of the methodologies across diverse wireless systems and applications requires thorough investigation. Furthermore, as these algorithms are mostly developed in simulation environments, comprehensive testing in practical settings is essential to facilitate real-time validation of their capability.

\section{Services with Mixed Criticality}\label{sec:mixed_criticality}
\subsection{From Network Slicing to Intent-Based Management}

Network slicing was introduced in 5G to support applications with different requirements in terms of
quality of service (e.g., transmission rate, latency, and reliability).
In a nutshell, a network slice is a logical network serving a certain business or customer need~\cite{zhang17-08a}. For example, one network slice can be set up to support mobile broadband applications with full mobility support and another slice can be set up to support IoT applications. These slices will all run on the same underlying physical core and radio networks and use the same transport network, but, from the end-user application perspective, they appear as independent networks, each with its unique capabilites. 
Network slicing relies on the 5G core network; this means that it requires the 5G network to operate in the so-called stand-alone mode. To date, however, most 5G networks operate in non-stand-alone mode, i.e., they connect to the 4G core network and rely on 4G for connection setup and mobility.
This should be seen as an intermediate step toward the stand-alone operation required to fully unleash the 5G capabilities.

Each network slice uses an appropriate set of network functions to realize the slice. In the radio-access network, slicing is not explicitly visible. Instead, packets to/from the core network are marked with a QoS Flow ID (QFI). Based on the QFI, the radio-access network process the data packets in the appropriate manner, for example prioritizing critical packets over non-critical packets in the scheduler or extending transmissions in time to achieve sufficient coverage for a coverage-critical massive IoT service.

Orchestrating and managing the different network slices can be a complex task~\cite{mei21-09a,wu22-02a}. It is expected that 6G will leverage the use ML
techniques to deploy \emph{intent-based network-management} solutions~\cite{leivadeas23-01a}. 
The underlying idea is that network management will be performed via an abstraction layer, which
will interpret and automatically translate the intents into network configurations  and, while operating the network, continuously learn and fine-tune the configurations based on the ongoing data communication. Given the recent advances in natural language processing, future systems may even allow the intents to be given in human language  (e.g., ``configure the network to enable $50$ IoT devices to share their local machine-learning models via federated learning, with a high differential-privacy requirement''). Intent-based network management is a powerful tool that could enable an operator to offer service-level agreements beyond the best-effort type of services provided today. For example, it could allow an operator to guarantee a certain data rate and latency characteristic  throughout a city as the intent-based management automatically and continuously monitors and adjusts the network configuration to meet this requirement, possibly by prioritizing this traffic over other traffic types such as massive IoT connections, which may have significantly relaxed latency requirements in comparison.

\subsection{Slicing of RAN Communication Resources: A Case Study}
We provide next a case study that highlights the network-slicing challenges and tradeoffs occurring
when services with mixed criticality need to be supported simultaneously.
For simplicity, we shall focus exclusively on the  handling of radio resources to meet the requirements of the different network slices. Let us assume that, together with the massive connectivity and critical services reviewed in
Sections~\ref{sec:massive_connectivity} and~\ref{sec:critical}, respectively, the network needs also to
support conventional mobile broadband traffic, which is characterized by large payload and stable
device-activation patterns.  

The simplest approach to this scenario is to allocate distinct and
non-overlapping radio resources in time and/or frequency to each one of these three services.
Following~\cite{popovski18-12f}, we shall refer to this approach as \emph{heterogeneous orthogonal
multiple access} (H-OMA).
It turns out that this approach is suboptimal and can be outperformed, in some scenarios, by a \emph{heterogeneous
non-orthogonal multiple access} (H-NOMA) approach where time-frequency resources are shared among
some of the services.
The key observation is that one can exploit the difference in the services requirements in terms of reliability 
via SIC decoding. 
In other words, one can exploit the inherent \emph{reliability diversity} among the services.

\begin{figure}[t]
  \centering
    \subcaptionbox{Broadband traffic\label{fig:embb}}[0.24\textwidth]{\includegraphics[width=0.24\textwidth]{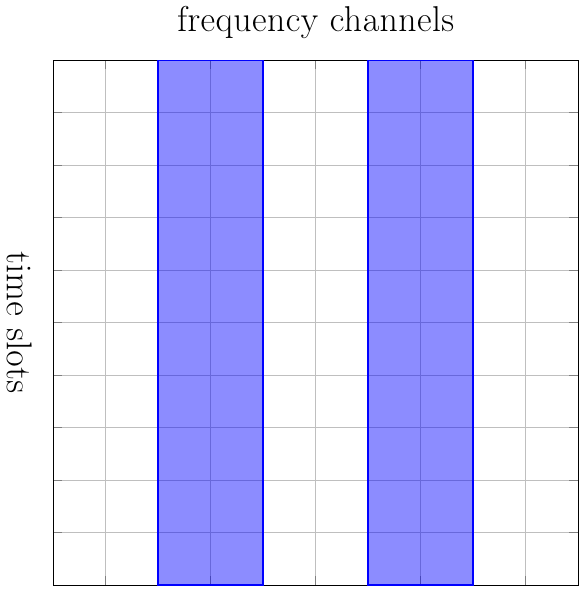}}
    \subcaptionbox{Critical traffic\label{fig:urllc}}[0.24\textwidth]{\includegraphics[width=0.24\textwidth]{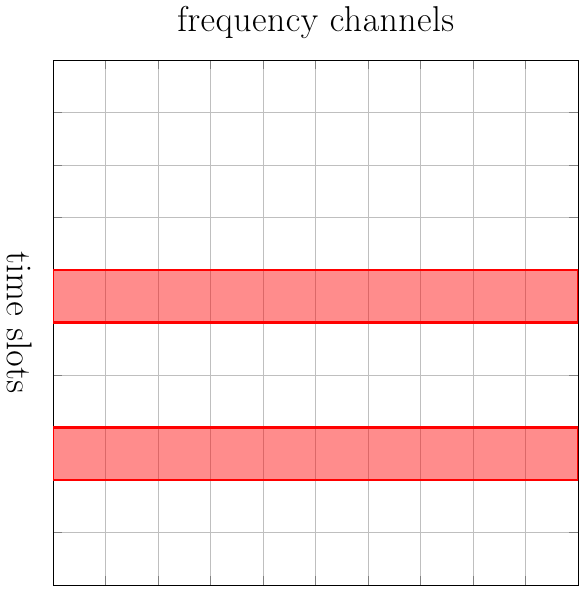}}
    \subcaptionbox{Massive
      traffic\label{fig:mmtc}}[0.24\textwidth]{\includegraphics[width=0.24\textwidth]{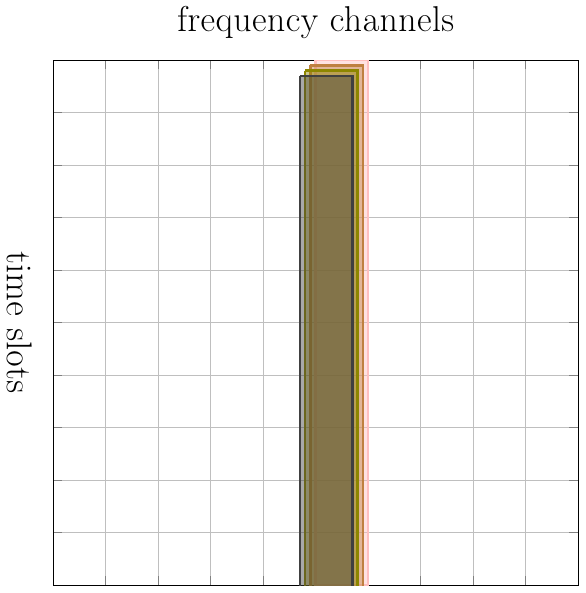}}
    \subcaptionbox{H-NOMA\label{fig:hnoma}}[0.24\textwidth]{\includegraphics[width=0.24\textwidth]{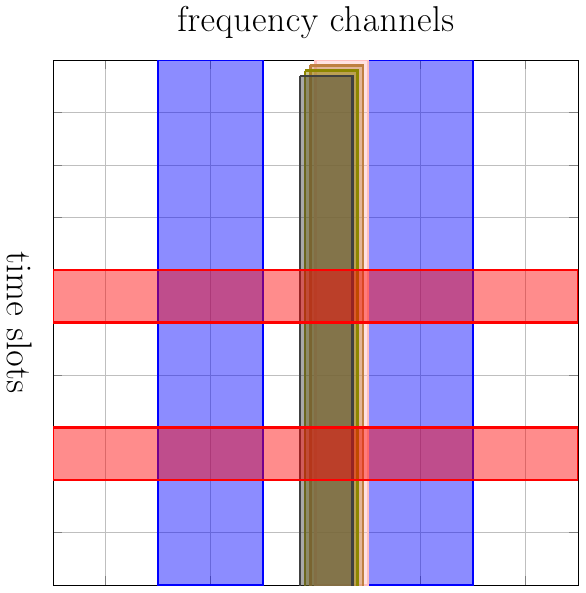}}
  \caption{(\subref{fig:embb})--(\subref{fig:mmtc}): Typical allocation of time-frequency resources for different kind of traffic typologies
  in 5G. (\subref{fig:hnoma}): Non-orthogonal slicing using H-NOMA.}\label{fig:traffic-patterns}
\end{figure}

To understand why this is the case, we will consider the problem of coexistence between broadband,
massive, and critical services as sketched in Fig.~\ref{fig:traffic-patterns}.
Let us first focus on the coexistence between broadband traffic
and critical services as originally defined in 5G.
From a communication theory perspective, we can model the broadband-traffic transmission problem as a
problem of communicating information over multiple frequency channels (see Fig.~\ref{fig:embb}) over a fading
channel that can be assumed to be known to both transmitter and receiver. 
This assumption is {a plausible approximation}, because of the large payloads that are typically assumed for broadband traffic, which make the cost of exchanging
channel-state information negligible. 
One reasonable objective when transmitting broadband traffic is to maximize the transmission rate
under  a long-term power constraint as well as a  constraint on the outage probability.

As proven in~\cite{caire99-07a}, the rate-maximizing strategy for this scenario is truncated channel
inversion,  according to which transmission occurs at a  power that is inversely proportional to
the square of the channel amplitude, provided that the channel gain is sufficiently large. 
If this is not the case, no transmission occurs, which results in an outage.

Furthermore, as illustrated in Fig.~\ref{fig:urllc}, we can model the critical-traffic transmission problem
as the problem of decoding the sporadic transmission of information allocated over a single time slot
of multiple frequency channels. 
Such traffic allocation has the advantage of minimizing the decoding delay.
For the transmission of critical traffic, it is reasonable to assume that no fading-channel knowledge is available at the
transmitter, since acquiring such a knowledge would typically result in large delays.
Also in this case, the outage probability for a given transmission rate is a natural performance
metric~\cite{yang14-07c}.

For this setup, H-OMA would entail reserving one or more time slots across a number of frequency
channels for critical traffic. 
As pointed out in Section~\ref{sec:intro}, though, this may be wasteful in terms of resources
because of the sporadic nature of critical-service transmission.
The approach in H-NOMA is to let both traffic types overlap in the same time-frequency resources, as illustrated in Fig.~\ref{fig:hnoma}.
The idea is to exploit the fact that critical services need to be decoded with a failure rate that
is typically $2$ to $3$ orders of magnitude smaller than broadband services. 
The receiver can exploit this reliability diversity by decoding the critical service first, and then
removing the interference it causes on the broadband service prior to its decoding. 
\begin{figure}[t]
  \centering
  \includegraphics[width=0.45\textwidth]{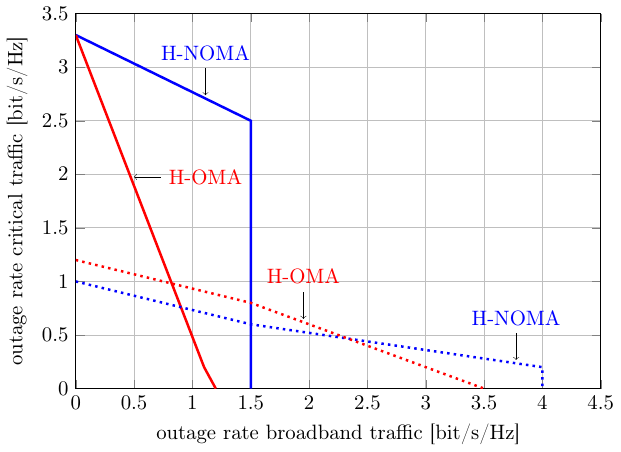}
  \caption{Pictorial representation of the tradeoff between outage rate of broadband and outage rate
    of critical
  traffic. 
  The solid lines correspond to the case in which the SNR experienced during the
transmission of the critical traffic is larger than that experienced during the transmission of the
broadband traffic. The dotted lines correspond to the opposite case.}\label{fig:embb-urllc-comp}
\end{figure}
As proven in~\cite{popovski18-12f} and shown pictorially in Fig.~\ref{fig:embb-urllc-comp}, this approach is
advantageous when the SNR experienced during the transmission of the critical traffic is larger than
the SNR experienced during the transmission of the broadband traffic.
This condition guarantees that the critical traffic can be decoded at the desired reliability level,
in the presence of the broadband-traffic interference, which is a necessary condition to enable interference
cancellation.
On the contrary, when the critical-traffic SNR is lower than the broadband-traffic SNR, H-OMA is
preferable. 

Next, we consider the coexistence between massive-connectivity and broadband traffic.
As illustrated in Fig.~\ref{fig:mmtc}, and in agreement with Section~\ref{sec:massive_connectivity}, we
can assume that this kind of traffic results in a random numbers of overlapping transmissions over a
pre-allocated frequency channel, each one occupying multiple time slots. 
As discussed in Section~\ref{sec:massive_connectivity}, a reasonable performance metric for this
kind of traffic is the fraction of correctly decoded transmissions.
A common approach in many massive random access protocols to address the interference
resulting from multiple transmission is SIC, where channel quality
and interference patterns are accounted for to successfully decode the active users. 

The H-NOMA approach in this context entails extending SIC, to
include the decoding of the broadband-traffic.
As shown in~\cite{popovski18-12f}, the order in which the data belonging to the two types of service
are decoded is critical from a performance perspective.
However, such ordering is difficult to determine analytically. 
A good heuristic is  to perform SIC decoding of the massive-access
users until no more users can be decoded. 
At this point, decoding of the broadband traffic is attempted, and, if it succeeds, the decoding of
the remaining massive-access users via SIC is resumed.
\begin{figure}[t]
  \centering
  \includegraphics[width=0.45\textwidth]{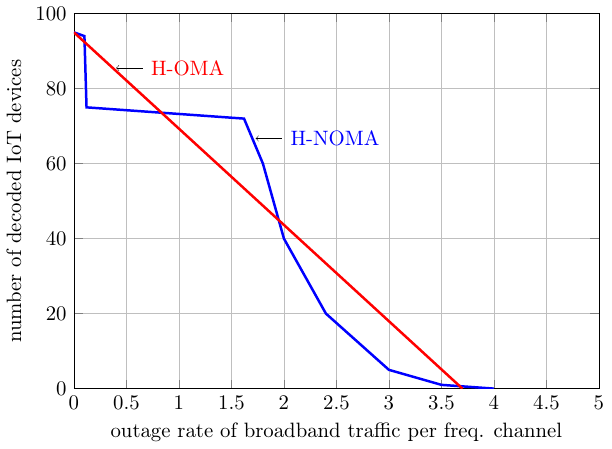}
  \caption{Pictorial representation of the tradeoff between outage rate of broadband traffic (per
  frequency channel) and average number of decoded IoT devices.}\label{fig:mmtc-broad-comp}
\end{figure}
As demonstrated in~\cite{popovski18-12f}, and illustrated pictorially in
Fig.~\ref{fig:mmtc-broad-comp}, the
H-NOMA approach is typically beneficial for intermediate values of the broadband-traffic rate,
i.e., when the broadband-traffic can be decoded only after few strong massive-connectivity users
have been decoded and their signal canceled. On the contrary, H-OMA is advantageous when the rate of
the broadband traffic is large.
Indeed, H-NOMA performs poorly in this regime because whenever the decoding of the broadband traffic
fails, so does the decoding of a large fraction of massive-connectivity users.

While no results are provided in~\cite{popovski18-12f} as far as slicing between critical and massive
connectivity is concerned, intuitively, for the resource-allocation model considered in Fig.~\ref{fig:traffic-patterns}, it seems unlikely that
one can satisfy with H-NOMA the reliability requirements of critical traffic, in the presence of the
random interference caused by massive connectivity.

\begin{figure}[t]
  \centering
  \includegraphics{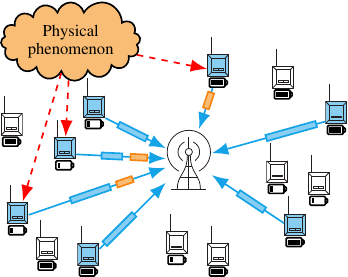}
  \caption{A 6G scenario for the coexistence of massive and critical connectivity. A large
  population of sensors may transmit both ordinary messages and alarm messages triggered by a physical phenomenon to a sink.
Alarm messages need to be decoded with much higher reliability than ordinary messages.}\label{fig:massive-critical}
\end{figure}

As pointed out in Section~\ref{sec:intro}, critical connectivity, latency and reliability do not
necessarily need to be coupled in 6G.
This suggests that the model for critical connectivity presented in Fig.~\ref{fig:urllc} should be
revised.
An alternative model is considered in~\cite{ngo23-10a}, where the authors assume that both
critical and massive services are provided over a number of consecutive time slots over the same
frequency channel, similar to what depicted in Fig.~\ref{fig:mmtc}. 
A possible relevant scenario for this setup is presented in Fig.~\ref{fig:massive-critical}: a large
population of sporadically active sensors transmit individual information to a sink. 
The occurrence of a certain physical phenomenon may, however, trigger a subset of sensors to
transmit an alarm message that needs to be decoded with much higher reliability than ordinary
messages. 

In such a setup, H-OMA would entail subdividing the slots available on the frequency channel into
slots dedicated to critical messages and slots dedicated to ordinary messages, whereas H-NOMA would
entail the transmission of a superposition of both messages when both are available.

For the simple physical-layer model considered in~\cite{ngo23-10a}, H-OMA turns out to be
advantageous over H-NOMA. 
Indeed, when using H-OMA, one can leverage the fact that, since all sensors conveying the alarm
message transmit the same codewords, such codewords combine coherently over the air, which implies
that only a small number of time slots is required to achieve the desired reliability. 
This makes the transmission of the alarm message very energy efficient.
Furthermore, even in situations where these slots are often idle because the alarm-triggering
physical phenomenon is rare, the ordinary traffic is not penalized, because the number of reserved
slots is small.
On the contrary, as pointed out already in~\cite{popovski18-12f}, H-NOMA suffers from the limited
performance of interference cancellation in such a scenario. Indeed, even when the alarm is
correctly decoded, it turns out to be difficult to estimate the number of sensors that transmitted
it to the level of accuracy required not to impact the decoding of the ordinary messages after SIC.
As a consequence, transmitting the alarm message with H-NOMA involves a much higher expenditure in
terms of energy per bit.

The network architecture described so far, in which only a single receiver is in charge of collecting
the signals from all users, may be replaced in 6G by a distributed architecture, where a multitude of
low-cost coordinated APs are deployed over a coverage area. 
As noted in, e.g.,~\cite{tominaga23-06a}, different AP deployments may be optimal for different
traffic typologies.
Going back to the example considered in Fig.~\ref{fig:massive-critical}, a uniform deployment of the
APs is typically optimal for the ordinary massive-connectivity traffic.
The situation is different in the alarm-message case. Indeed, if we require the alarm message from
each sensor to be decoded individually, then a centralized deployment, where the distributed APs are
replaced by a centralized massive antenna array, results typically in better performance.
Indeed, the locality of the physical event generating the alarm traffic, implies that the locations
of the activated sensors are highly correlated.
As a consequence, a receiver with a large number of antennas is required to 
separate their signals.

\subsection{Network Slicing for Decentralized Learning and Inference: Challenges and Open Problems}
As pointed out in Section~\ref{sec:intro}, the main driver for massive connectivity in 6G is
expected to be distributed/decentralized learning and inference. 
This raises the question of whether this kind of traffic should be allocated a separate
network slice and, if so, which kind of key performance indicators should be associated to this
network slice.
This is an active area of research and no accepted metric is available so far.
Obviously, one can choose a specific distributed learning application, as often done in the
literature, and exhibit results in terms of training convergence or generalization error.
However, such an approach lacks generality.

Another interesting challenge brought by this traffic typology is how to exploit effectively the inherent
communication-computation convergence opportunities.
Such convergence opportunities arise from the fact that, in distributed/decentralized learning and
inference applications, the sink is often
interested only in a function of the signals transmitted by the active users.
For example, in the case of over-the-air federated learning---one of the applications that have been
studied more in depth in the literature~\cite{amiri20-05a,amiri21-08a}---each active user may send a
quantized version of its
local gradient, whereas the sink is interested only in the arithmetic average of the local
gradients.
The initial approaches proposed in the literature to perform such an aggregation over the air, rely on the transmission of
analog waveforms that are not compatible with the digital communication protocols used in cellular
communication standards.
Digital-domain versions of such initial approaches have been recently investigated in,
e.g.,~\cite{qiao23-05a}.
Such recent protocols rely on a combination of the grant-free massive access approach
proposed~\cite{polyanskiy17ura} with the type-based multiple-access approach proposed in
e.g.,~\cite{mergen06-02a}.
This promising combination augments the framework proposed in~\cite{polyanskiy17ura} with an explicit
mapping between messages and observations, and with the requirement that the multiplicity of each
observation is also detected by the decoder~\cite{gunduz23bits}.
However, no fundamental bounds on the performance achievable using this approach are available, which
makes it difficult to assess the optimality of the protocols proposed so far.

\section{Concluding remarks}\label{sec:conclusion}

The focus on massive communication and critical services started in 5G, with the consideration of mMTC and URLLC as two generic services. This article has discussed the evolution of massive and critical connectivity towards 6G, reflecting the updated understanding of the requirements for these services as well as the methods and tools available to design and assess those services. Regarding massive connectivity, besides the protocols and algorithms for uplink access relevant already in 5G, we have presented methods for downlink transmission, necessary to support massive closed loop communication. The discussion on critical services was centered on the generalization of latency-reliability requirements towards other types of timing requirements. The focus was also put on models of channel and traffic dynamics for attaining the critical service requirements. Notably, we have discussed the role of AI technologies in meeting and assessing the requirements for criticality and robustness. Finally, we have discussed the problem of sharing wireless resources among services with different criticality. This part is an important illustration of the existence of use cases in which massive and critical communication converge.


\end{document}